# The New CMS Measure of Excessive Radiation Dose or Inadequate CT Image Quality: Methods for Size-Adjusted Dose and Their Variabilities

Gary Y Ge, PhD, Charles M Weaver, MS, Jie Zhang, PhD

Division of Diagnostic and Nuclear Medical Physics, Department of Radiology, University of Kentucky College of Medicine, Lexington, KY 40536

## Introduction

The Centers for Medicare & Medicaid Services (CMS) has introduced CMS1074v2, a new CT quality measure aimed at limiting excessive radiation dose while maintaining image quality. The measure evaluates the percentage of studies exceeding thresholds for image noise (poor quality) or radiation dose (excessive exposure) across 18 CT exam categories [1]. It has been incorporated into major CMS quality-based payment programs, impacting hospital and clinician reimbursements. Reporting began in January 2025, with mandatory compliance from 2027.

Following its initial development, a wide spectrum of concerns has been raised regarding the measure's definitions, methodological transparency, terminology, and alignment with clinical standards. Compliance requires calculating three key variables per CT study: CT Dose and Image Quality Category, Calculated CT Global Noise (GN), and Calculated CT Size-Adjusted Dose (SAD). For SAD, the effective diameter is compared against the UCSF International CT Dose Registry median diameter for the respective CT Dose and Image Quality Category and modified by a category-specific size-adjustment coefficient. This approach has significant limitations in characterizing certain body regions, such as the thorax, and deviates from the IEC methodology endorsed by the FDA, raising concerns about consistency in follow-up CT dosimetry across scanners, institutions, and study types.

This study focuses on SAD, examining its calculation methods and potential variability in thoracic and abdominal CT exams. Variability in SAD determination may compromise the measure's goal of optimizing radiation dose while preserving diagnostic integrity.

## SAD Calculation

The method for calculating SAD provided by CMS is as follows [2]:

Eq 1. $$D_A = D_R e^{-\beta_k(d-d_k)}$$

Where $D_A$ is the SAD value, $D_R$ is the unadjusted DLP, $d$ is the diameter of the anatomic area being examined, $d_k$ is the "expected diameter" of the CT category associated with the exam, and $\beta_k$ is the "size-adjustment coefficient" of the CT category associated with the exam. The values for $d_k$ and $\beta_k$ are derived from the UCSF International CT Dose Registry and specified in the measure for each of the 18 CT categories [2, 3].

While DLP is directly available from each CT examination, the diameter of the anatomic area (d) must be measured or estimated. Effective diameter can be computed using AP and/or lateral measurements from the axial image [4]. If axial images are unavailable, it may be derived from the localizer image [5]. Currently, five methods are available for estimating effective diameter based on different scenarios: HU Value Thresholding, Water-Equivalent Diameter (WED), LAT or AP Conversion Method, $\sqrt{LAT \times AP}$ Method, and (LAT + AP) / 2 Method [4-6].

Since CMS does not mandate a specific method for estimating effective diameter, all five methods were tested on the collected patient image data. Three metrics were calculated, $d$, SAD and $D_A/D_R$. $D_A/D_R$ represents the exponential scaling factor that results from the difference between $d$ and $d_k$, thus it should provide DLP-independent insight into how the patient size calculation methods operate on the patient cohort.

For the AP/LAT Conversion method, LAT thickness was measured on the scout image as specified in TG204. However, for the (LAT + AP)/2 and $\sqrt{LAT \times AP}$ methods, AP and LAT thicknesses were measured directly on the axial middle slice, as our institution does not routinely acquire two scout views. Additionally, because the Savitzky-Golay smoothing filter used in Christianson 2012 lacked defined parameters, a 15-pixel window was applied, as it demonstrated consistent performance across the patient cohort.

## SAD Evaluation

*Patient Cohort Specifications*

Patient images for seven CT protocols were retrospectively retrieved for SAD evaluation. These included five abdomen and two chest protocols and covered a range of CT exam categories based on examples outlined in the UCSF study referenced in the CMS measure, as shown in Table 1 [1, 2, 7].

Table 1. Selected CT protocols and associated body region and dose categories.

| CT Protocol | Body Region | CMS Dose Category |
|---|---|---|
| Renal Stone | Abdomen | Low |
| Abdomen Pelvis w/o | Abdomen | Routine |
| Enterography | Abdomen | Routine |
| Urogram | Abdomen | High |
| Renal Mass w/o | Abdomen | High |

| Chest w/o IV Contrast | Chest | Routine |
|---|---|---|
| Chest PE | Chest | Routine |

The selection criteria for the exams were as follows: no anatomical cutoff in the scan midslice, no implanted medical devices present in the scan midslice, no arms down in the imaging plane. These criteria ensured that all patient size methods could be calculated and that there were no high-attenuation materials that could affect patient size calculations.

In total 719 exams were collected for the study cohort, with at least 100 exams per CT protocol. A demographic breakdown can be seen in Table 2.

Table 2. Patient cohort characteristics

| Characteristics | Number of patients | Percentage of patients (%) |
|---|---|---|
| *n*, number of patients | 719 | NA |
| Gender | | |
| Male | 358 | 51.4 |
| Female | 361 | 48.6 |
| Age | | |
| 18-29 | 88 | 12.2 |
| 30-39 | 75 | 10.4 |
| 40-49 | 93 | 12.9 |
| 50-59 | 144 | 20.0 |
| 60-69 | 183 | 25.5 |
| 70-79 | 105 | 14.6 |
| 80-89 | 29 | 4.0 |
| 90-99 | 2 | 0.3 |
| BMI | | |
| Mean | 27.6 | |
| SD | 6.0 | |
| 50th | 26.9 | |
| 75th | 31.3 | |

*Statistical Analysis*

For each protocol, the SAD values for each patient size calculation method were compared to the SAD thresholds outlined in the CMS measure. Kruskal-Wallis testing was performed to determine significant differences for $d$, SAD, and $D_A/D_R$ across all size calculation methods, respectively ($p < 0.05$). Pairwise comparisons were performed to determine which methods were significantly different for each metric ($p < 0.05$). Statistical analysis was performed using MATLAB 2024a (MathWorks, Natick, MA, USA).

**Results**

Table 3 shows effective diameter $d$, SAD calculations based on five available methods for all CT protocols. It also includes the number of exams that exceed the CMS SAD threshold for each patient size calculation method. Notably, a large portion of the Chest PE protocol exceeds the threshold value for routine chest protocols (377 mGy.cm). For the chest protocols, the attenuation-based methods (HU threshold and WED) cause more exams to fail than the lateral and AP based methods. This trend appears to be reversed in abdominal protocols, though the impact is both limited and inconsistent. SAD values for the Urogram CT protocol are well under the CMS threshold value, despite being listed as a high dose abdomen protocol.

Figure 1 shows the relative effect on SAD when the calculated patient size differs from the "expected diameter", as defined by the CMS regulation for each dose category. If the calculated diameter is smaller than the expected diameter, SAD is overestimated. Conversely, if the calculated diameter is larger, SAD is underestimated.

Figure 2 compiles boxplots of SAD for all included CT protocols, showing a chart outlining the mean SAD for each CT protocol, side-by-side. The SAD boxplots suggest that the attenuation-based methods (HU threshold and WED) tend to overestimate the patient diameter relative to the projection-based methods in the abdomen protocols, and vice versa in the chest protocols.

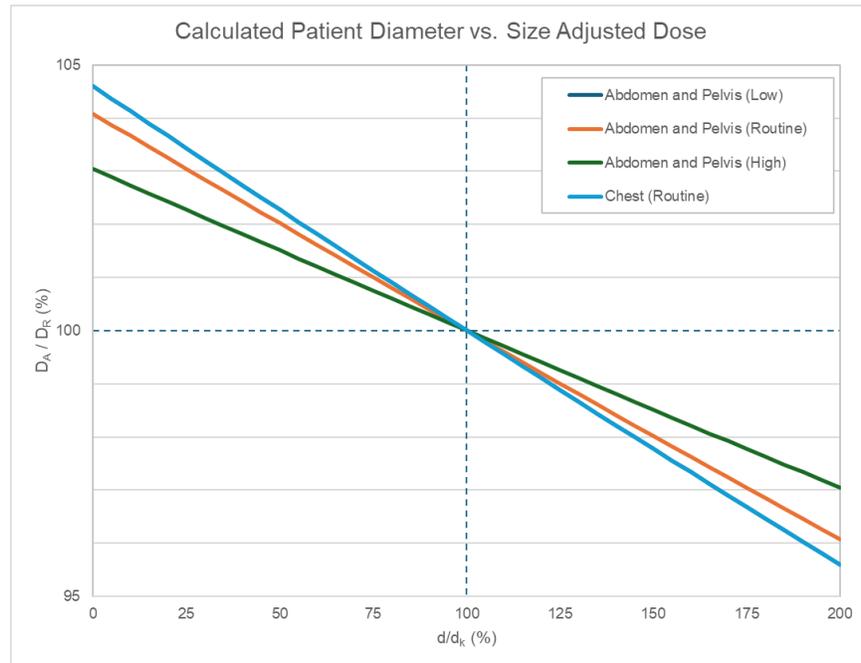

Figure 1. Impact on size-adjusted DLP value due to difference in calculated vs expected diameter. When the calculated diameter is smaller than the expected diameter, the size-adjusted DLP is overestimated, and vice versa. Each CMS defined dose category has an assigned $\beta_k$ value, which determines the slope of this relationship. Abdomen and Pelvis (Low) and Chest (Routine) have the same $\beta_k$ values, so the lines overlap in this chart.

Table 3. General summary of calculations for included CT protocols.

| CT Protocol | CMS Category | # of Exams | Metrics | CMS HU Threshold | WED | LAT Conversion | √(LAT X AP) | (LAT + AP)/2 |
|---|---|---|---|---|---|---|---|---|
| **Renal Stone** | Abdomen (Low) | 100 | d | 306.0 (43.6) | 299.3 (38.4) | 292.7 (57.0) | 291.5 (40.1) | 295.4 (39.5) |
| | | | SAD | 351.2 (101.8) | 372.7 (107.2) | 395.7 (106.9) | 400.1 (115.3) | 386.1 (109.7) |
| | | | # Exceeding CMS | 3 | 3 | 1 | 7 | 6 |
| **Abd Pelvis WO** | Abdomen (Routine) | 100 | d | 303.2 (42.2) | 298.2 (37.7) | 305.2 (59.7) | 290.5 (38.6) | 294.7 (38.3) |
| | | | SAD | 418.1 (143.7) | 436.8 (155.7) | 412.5 (139.5) | 461.4 (153.5) | 445.4 (145.2) |
| | | | # Exceeding CMS | 11 | 12 | 6 | 13 | 11 |
| **Enterography** | Abdomen (Routine) | 106 | d | 299.3 (44.7) | 297.3 (38.3) | 291.1 (55.5) | 284.8 (42.5) | 288.8 (42.5) |
| | | | SAD | 334.3 (126.6) | 343.3 (140.0) | 359.4 (128.9) | 376.4 (145.3) | 363.8 (138.4) |
| | | | # Exceeding CMS | 3 | 5 | 3 | 5 | 4 |
| **Urogram** | Abdomen (High) | 100 | d | 308.4 (37.9) | 302.8 (34.0) | 303.8 (46.9) | 293.9 (34.9) | 297.4 (34.2) |
| | | | SAD | 351.5 (164.1) | 365.3 (174.3) | 349.1 (136.7) | 384.7 (182.3) | 375.9 (177.2) |
| | | | # Exceeding CMS | 0 | 0 | 0 | 0 | 0 |
| **Renal Mass WO** | Abdomen (High) | 105 | d | 333.5 (46.6) | 327.4 (42.1) | 331.7 (60.3) | 316.6 (42.7) | 318.5 (42.3) |
| | | | SAD | 757.6 (293.5) | 792.5 (321.6) | 762.0 (277.2) | 844.0 (341.6) | 834.3 (337.1) |
| | | | # Exceeding CMS | 5 | 6 | 5 | 10 | 9 |
| **Chest WO** | Chest (Routine) | 105 | d | 262.2 (39.4) | 263.9 (38.4) | 284.2 (56.2) | 301.8 (32.7) | 304.8 (33.1) |
| | | | SAD | 287.5 (157.0) | 284.4 (157.5) | 246.9 (166.0) | 199.8 (113.5) | 194.3 (109.8) |
| | | | # Exceeding CMS | 28 | 24 | 17 | 4 | 4 |
| **Chest PE** | Chest (Routine) | 107 | d | 256.4 (37.3) | 265.3 (36.9) | 272.6 (50.6) | 288.0 (29.0) | 291.5 (29.3) |
| | | | SAD | 466.6 (435.7) | 433.6 (425.7) | 431.3 (633.5) | 349.3 (350.5) | 336.6 (326.5) |
| | | | # Exceeding CMS | 43 | 40 | 33 | 25 | 22 |

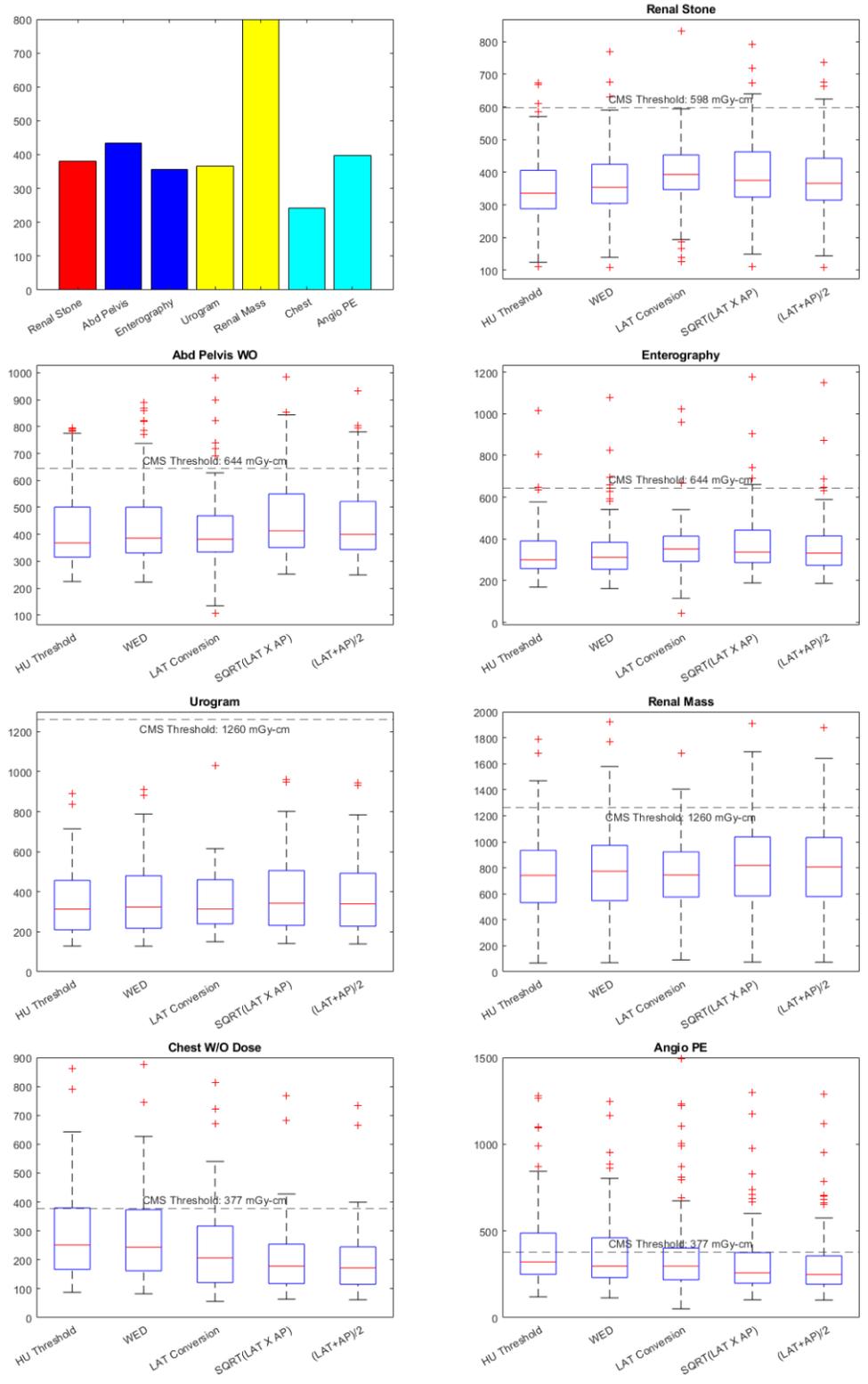

Figure 2. SAD calculations for each of the evaluated CT protocols. Five different patient diameter calculation methods are used, which covers all methods that are supported by citations within the CMS measure. A bar plot showing the mean SAD values for each protocol is also included.

The Kruskal-Wallis analysis indicates significant difference between calculation methods for several protocols and calculated metrics ($d$, SAD, $D_A/D_R$). All protocols, except Urogram and Abd/Pelvis WO, exhibit significant differences across groups. Pairwise evaluation did not show any significant differences within Abd/Pelvis WO. Figure 3 presents the pairwise evaluations for all CT protocols and calculated metrics, with red-highlighted boxes indicating significant differences due to the choice of calculation methods.

Statistical differences in $d$ and $D_A/D_R$ calculations are associated with the different patient size calculation methods, since they are both independent of individual exam DLP values. On the other hand, statistical differences in SAD have dependence on the patient cohort and institutional protocols. The results suggest that the calculation methods are inconsistent for the Urogram, Chest WO and Chest PE protocols, as these have more statistical differences in $d$ and $D_A/D_R$.

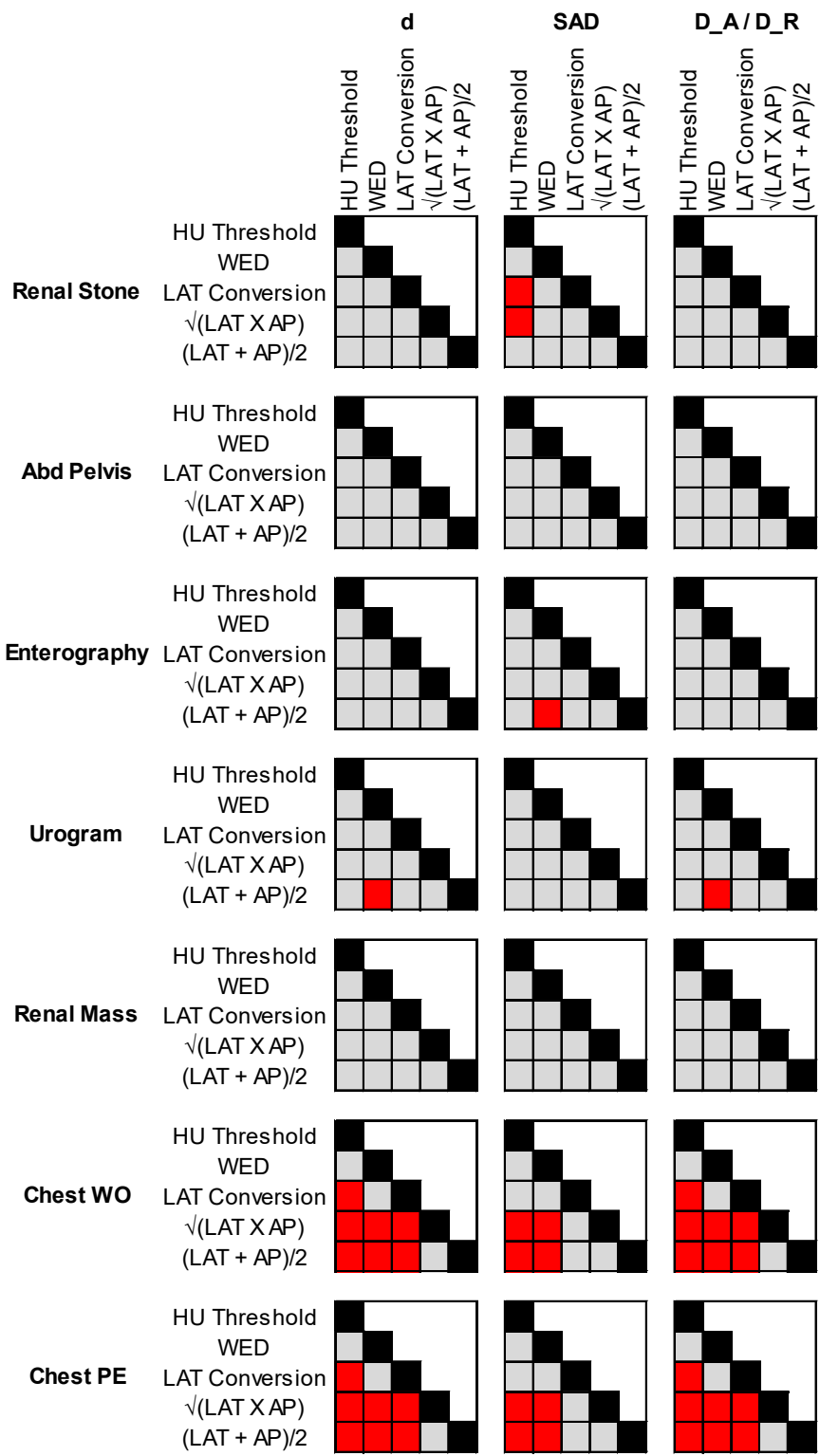

Figure 3. Comparisons showing significant differences between pairwise comparisons for all CT protocols and calculation methods. The parameters compared are patient diameter (d), size-adjusted DLP (SAD), and the exponential scaling applied to DLP to produce SAD ($D_A/D_R$).

## Discussion

This study examines methods for SAD calculation and potential variations. Preliminary testing revealed that approximately 30% of CT exams failed one or both measure metrics, with SAD accounting for over 99% of the failures, highlighting its significant impact on institutional compliance [8]. The measure's language lacks specificity in defining dose metric calculations, leading to inconsistencies. Statistical analysis shows no universally optimal method for patient size calculation, as results depend on the anatomical region, limiting the CMS measure to consistently assess a broad range of CT protocols.

Concerns also arise regarding the accurate classification of CT protocols. The Urogram protocol examined in this study should fall under the high-dose abdomen category, based on a cited source within the measure. However, SAD values at our institution are more than 30% below the defined threshold, suggesting possible misclassification, either locally or within the measure itself. The reliance on machine-stated protocols rather than CPT codes or other metadata introduces further uncertainty, particularly given institutional variations in how these attributes are assigned.

Beyond individual protocols, CT exam categorization is influenced by broader metadata practices. The use of DICOM tags for dose tracking and reporting is a key component, yet institutional differences in metadata conventions can lead to classification errors. A UCSF study found that even under controlled conditions, only 90% of randomly selected data were categorized correctly [7]. In institutions with differing DICOM metadata conventions, classification accuracy could be even lower. The lack of a publicly available, standardized method for determining CT dose categories presents a significant challenge for ensuring consistency across institutions.

This study has limitations, primarily related to cohort size and CT protocol selection. The cohort was selected to allow all calculation methods to be performed on each patient image series, favoring smaller patients and making the sample unrepresentative of the local population. For example, in a one-month search period for Chest PE protocol, only 67 out of 239 studies (28%) met inclusion criteria, as most were excluded due to anatomic cutoff in the midslice. This suggests that topogram-based methods will be heavily relied upon in clinical implementation. However, these methods demonstrated the most inconsistency in our results, raising additional concerns about the reliability of the CMS measure.

In conclusion, our findings reinforce the concerns highlighted by the AAPM Commissioned Panel [9]. While CMS regulations represent progress toward standardized radiation safety practices, ambiguities in technical implementation raise questions about the effectiveness of the proposed metrics. The flexibility in methodology, though beneficial in some cases, may compromise consistency and comparability of reported dose metrics. Variability in institutional interpretations could undermine the intent of regulatory standardization. Addressing these uncertainties will require ongoing collaboration and refinement of methodologies to ensure that radiation dose reporting remains accurate, meaningful, and actionable for patient safety and clinical decision-making.

# Citations

1. *Excessive Radiation Dose or Inadequate Image Quality for Diagnostic Computed Tomography (CT) in Adults*. 2024, Centers for Medicare and Medicaid Services.
2. *Excessive Radiation Dose or Inadequate Image Quality for Diagnostic Computed Tomography (CT) in Adults (Clinician Level)*. 2022, National Quality Forum.
3. Smith-Bindman, R., et al., *Radiation Doses in Consecutive CT Examinations from Five University of California Medical Centers.* Radiology, 2015. **277**(1): p. 134-41.
4. Cheng, P.M., *Automated estimation of abdominal effective diameter for body size normalization of CT dose.* J Digit Imaging, 2013. **26**(3): p. 406-11.
5. Christianson, O., et al., *Automated size-specific CT dose monitoring program: assessing variability in CT dose.* Med Phys, 2012. **39**(11): p. 7131-9.
6. AAPM, *Use of Water Equivalent Diameter for Calculating Patient Size and Size-Specific Dose Estimates (SSDE) in CT*. 2014.
7. Smith-Bindman, R., et al., *An Image Quality-informed Framework for CT Characterization.* Radiology, 2022. **302**(2): p. 380-389.
8. *CT Quality & Safety: Measure Testing*. [cited 2024 November 6]; Available from: https://ctqualitymeasure.ucsf.edu/measure-details/measure-testing.
9. Wells, J.R., et al., *The New CMS Measure of Excessive Radiation Dose or Inadequate Image Quality in CT: Issues and Ambiguities—Perspectives from an AAPM-Commissioned Panel.* American Journal of Roentgenology. **0**(ja): p. null.